\documentclass[final,twocolumn,pre,aps,amsfonts,amssymb,showpacs,superscriptaddress,floatfix]{revtex4-1}
\usepackage{enumerate}
\usepackage{amsmath}
\usepackage{graphicx}
\usepackage{color}
\usepackage{hyperref}
\usepackage{psfrag}
\usepackage{stmaryrd}
\usepackage{amssymb}
\usepackage{wasysym}
\usepackage{soul}

\def \degree {^\mathrm{o}}

\begin{document}

\title{Intermittent stick-slip dynamics during the peeling of an adhesive tape from a roller}
\author{Pierre-Philippe Cortet}
\affiliation{Laboratoire FAST, CNRS, Univ Paris Sud, UPMC Univ
Paris 06, France}
\author{Marie-Julie Dalbe}
\affiliation{Institut Lumi\`ere Mati\`ere, UMR5306 Universit\'e Lyon 1-CNRS, Universit\'e de Lyon, France}
\affiliation{Laboratoire de Physique de l'Ecole Normale Sup\'{e}rieure de Lyon, CNRS and Universit\'{e} de Lyon, France}
\author{Claudia Guerra}
\affiliation{Institut Lumi\`ere Mati\`ere, UMR5306 Universit\'e Lyon 1-CNRS, Universit\'e de Lyon, France}
\author{Caroline Cohen}
\affiliation{Laboratoire de Physique de l'Ecole Normale
Sup\'{e}rieure de Lyon, CNRS and Universit\'{e} de Lyon, France}
\author{Matteo Ciccotti}
\affiliation{Laboratoire PPMD/SIMM, UMR7615 (CNRS, UPMC, ESPCI
Paristech), Paris, France}
\author{St\'{e}phane Santucci}
\affiliation{Laboratoire de Physique de l'Ecole Normale
Sup\'{e}rieure de Lyon, CNRS and Universit\'{e} de Lyon, France}
\author{Lo\"{i}c Vanel}
\affiliation{Institut Lumi\`ere Mati\`ere, UMR5306 Universit\'e Lyon 1-CNRS, Universit\'e de Lyon, France}

\date{\today}

\pacs{62.20.mm
, 68.35.Np
, 82.35.Gh
}

\begin{abstract}
We study experimentally the fracture dynamics during the peeling
at a constant velocity of a roller adhesive tape mounted on a
freely rotating pulley. Thanks to a high speed camera, we measure,
in an intermediate range of peeling velocities, high frequency
oscillations between phases of slow and rapid propagation of the
peeling fracture. This so-called stick-slip regime is well known
as the consequence of a decreasing fracture energy of the adhesive
in a certain range of peeling velocity coupled to the elasticity
of the peeled tape. Simultaneously with stick-slip, we observe low
frequency oscillations of the adhesive roller angular velocity
which are the consequence of a pendular instability of the roller
submitted to the peeling force. The stick-slip dynamics is shown
to become intermittent due to these slow pendular oscillations
which produce a quasi-static oscillation of the peeling angle
while keeping constant the peeling fracture velocity
(averaged over each stick-slip cycle). The observed correlation
between the mean peeling angle
and the stick-slip amplitude questions the validity of the usually
admitted independence with the peeling angle of the fracture
energy of adhesives.
\end{abstract}

\maketitle

\section{Introduction}
The stick-slip instability that can develop during the high speed
peeling of adhesives, and which consists in strong oscillations
between phases of slow and rapid propagation of the peeling
fracture, constitutes a major problem in the polymer industry. The
scratchy sound that anyone can experience when pulling on an
adhesive tape, which is a trace of this instability, can indeed
cause a level of acoustic noise that is simply unbearable in the
industrial context. Another negative impact of stick-slip is the
damage caused to the adhesive
coating~\cite{Ryschenkow1996,Yamazaki2006} when the instability
occurs during the peeling of a temporary substrate layer before
the adhesive is effectively used. It is for example a severe
problem for hard disk drive (HDD) manufacturers as stick-slip will
deteriorate the quality of the adhesive seal which can lead to HDD
failure. These industrial concerns have recently conducted many
patents on this issue to be deposited (e.g.~\cite{Nonaka2009}).
Overall, adhesive stick-slip reduces industrial productivity and
its current hard-to-predict nature hinders the development of new
technical applications.

From a fundamental perspective, this unstable stick-slip crack
growth is admitted to be the consequence of a decreasing fracture
energy $\Gamma(v_p)$ in a certain range of peeling fracture
velocity $v_p$. This anomalous drop of the fracture energy has
been proposed to be related to structural transitions, from
cohesive to interfacial failure~\cite{Gent1969}, or between
different interfacial failure modes~\cite{Derail1997}. It has
however also been proposed~\cite{deGennes1996} that the
rheological transition of adhesive materials ---from soft to hard
rubber or from rubber to glass--- as a function of the strain rate
could be, in the presence of confinement (which is the case for
adhesive tapes), at the origin of a drop in the cohesive fracture
energy. Overall, the stick-slip motion, resulting from this
decreasing zone of fracture energy coupled to the compliance of
the peeled tape or peeling machine, corresponds to an oscillation
of the crack velocity between two (usually) very different values.
There are several factors that may influence the peeling velocity
range in which stick-slip effectively appears. For instance, the
stick-slip velocity thresholds can show a dependence on the glass
transition temperature of the
adhesive~\cite{Aubrey1980,Derail1997}, the thickness of the
adhesive layer~\cite{Gardon1963,Kim1989}, the substrate
roughness~\cite{Racich1975} and its viscoelastic
properties~\cite{Renvoise2007}. Remarkably, when stick-slip
occurs, the details of its dynamics change with the imposed
peeling velocity but also with the length of the tape submitted to
the peeling load~\cite{Barquins1986} and sometimes the stiffness
of the loading machine~\cite{Yamazaki2006}.

As proposed and verified experimentally by
Kendall~\cite{Kendall1975}, the fracture energy of a peeled
adhesive tape does not depend on the peeling angle in the regular
and slow (with respect to the stick-slip domain) peeling regime,
which result is widely extrapolated to larger peeling velocities.
An effect of the peeling angle on the velocity range for which
stick-slip exists was nevertheless already reported in some early
experiments~\cite{Aubrey1969}, however in conditions where the
length of the peeled tape was not constant but instead linearly
increasing with time during the peeling.

In this paper, we describe experiments of adhesive tape peeling
from a freely rotating roller in which we aim at imposing the
peeling velocity and the peeled tape length, defined as the
distance between the peeling fracture front on the roller and a
winding cylinder. Keeping these two parameters constant is indeed
necessary to produce a well-defined stick-slip
dynamics~\cite{Barquins1986}. Thanks to a fast imaging camera
coupled to image correlation velocimetry, we are able to extract
the full dynamics of the peeling fracture velocity with respect to
the substrate. In practice, we do not impose the peeled tape
length but only the distance between the adhesive roller and the
winding cylinder (Fig.~\ref{fig:setup}). During an experiment at
constant pulling velocity, superimposed on the stick-slip
instability, we may observe a slow oscillation of the angular
position at which the tape pulls on the roller. This slow dynamics
causes the effective peeling angle (averaged over one stick-slip
event) to oscillate with significant amplitude but in a
quasistatic manner for the stick-slip. We report that the value of
the effective peeling angle has a strong effect on the triggering
and amplitude of the stick-slip instability, even tough the mean
fracture velocity and peeled tape length remain constant or at
least not significantly affected by the slow oscillations. This
effect of the peeling angle on stick-slip cannot be simply
understood by taking into account its influence on the work term
of the elastic energy release rate as proposed by
Kendall~\cite{Kendall1975}. We suggest that the detailed features
of any adhesive stick-slip motion should depend not only on the
peeling velocity and peeled tape stiffness, but also strongly on
the effective peeling angle.

\section{Experimental setup}
\begin{figure}
\centerline{\includegraphics[width=8.5cm]{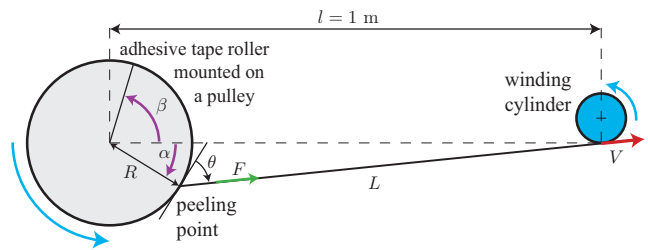}}
    \caption{(Color online) Schematic view of the experimental setup. The angles
    $\alpha$ and $\beta$ are oriented \emph{clockwise} and
    \emph{counterclockwise} respectively. Roller diameter:
    $40$~mm$<2R<58$~mm, roller and tape width: $b=19$~mm, tape
    thickness: $e=58~\mu$m.}\label{fig:setup}
\end{figure}

We peel a roller adhesive tape, mounted on a freely rotating
pulley, by winding up the peeled ribbon extremity on a cylinder at
a constant linear velocity $V$ using a servo-controlled brushless
motor (Fig.~\ref{fig:setup}). The distance between the pulley and
the winding cylinder is fixed to $l=1$~m. It is defined between
the adhesive roller center and the point, assumed to be fixed, at
which the peeled tape joins the winding spool. The adhesive tape
used, 3M Scotch$^\circledR$ 600, of the same kind as in
Refs.~\cite{Barquins1997,Cortet2007}, is made of a polyolefin
blend backing ($38~\mu$m thick) coated with a $20~\mu$m layer of a
synthetic acrylic adhesive. Each experiment consists in increasing
the winding velocity from $0$ up to the target velocity $V$ at a
rate of $1$~m~s$^{-2}$. Once the peeling velocity $V$ is reached,
it is maintained constant to a precision better than $\pm 2\%$
during two seconds, before decelerating back to zero. We have
varied the imposed velocity $V$ from $0.15$ to $2.55$~m~s$^{-1}$
in order to cover the whole range where stick-slip instability is
observed for the considered adhesive tape and peeling geometry.

The local dynamics of the peeling fracture line, viewed as a point
from the side, is imaged using a high speed camera (Photron Ultima
1024) at a rate of $f=8\,000$~fps and a resolution of $512 \times
64$~pixels. The field of view being approximately 2.5~cm wide, the
resolution is about $50~\mu$m/pixel. The recording of each movie
is triggered once the peeling has reached a constant average
velocity $V$ in order to obtain a stationary condition for the
peeling experiment. Following the method presented in
\cite{Cortet2007}, correlations between images of the movie,
separated of a time $\delta t=N/f$ ($N\in \mathbb{N}$), allow to
access:
\begin{itemize}
\item the curvilinear position of the peeling point in the
laboratory reference frame $\ell_\alpha = R\, \alpha $, where
$\alpha$ is the angular position of the peeling point (chosen
positive in the \emph{clockwise} direction, $\alpha>0$ in
Fig.~\ref{fig:setup}) and $R$ is the roller radius (between 20 and
29~mm), \item and, the curvilinear position of the adhesive roller
$\ell_\beta =R\,\beta$, in the laboratory reference frame, where
$\beta$ is the unwrapped angular position of the roller (chosen
positive in the \emph{counterclockwise} direction, $\beta>0$ in
Fig.~\ref{fig:setup}).
\end{itemize}
We are finally able to compute the curvilinear position $\ell_p$
of the peeling fracture point in the roller reference frame
($\ell_p$ is chosen so that it increases when the peeling front
advances) \begin{equation}
\ell_p =\ell_\alpha+\ell_\beta= R (\alpha + \beta). \label{eq.lp}
\end{equation}
We can then compute the peeling fracture velocity $v_p$ relative
to the substrate
\begin{equation}
v_p=\frac{d\ell_p}{dt}=R (\dot{\alpha}+\dot{\beta}). \label{eq.vp}
\end{equation}
Here, the substrate simply consists in the backing of the adhesive
tape remaining to peel.

The separation number $N$ between the images used for correlation
is chosen such that the moving matter at the periphery of the
roller displaces of about 5 pixels ($\sim250~\mu$m) between the
two images. Since the correlation is subpixel interpolated, we
reach a precision of about 1~pixel$/10\sim 5~\mu$m on the
displacement, i.e. 2\%. We finally get the same precision of 2\%
on the average peeling point velocity $v_p$ over a timescale $dt
\sim (250\times 10^{-6}$~m$)/V$, varying between 1.7~ms at the
lowest imposed velocity and down to 0.1~ms at the largest imposed
velocity.

\section{Equations of motion}
The equation ruling the motion of the adhesive roller can be
written as
\begin{equation}
I \ddot{\beta} = F R\,\cos \theta,
\label{eq.roller}
\end{equation}
where $I$ is the moment of inertia of the roller and $F$ the
tensile force transmitted along the peeled tape. Here, the angle
$\theta$ and $\alpha$ are linked by the geometrical constraint
\begin{equation}
l \cos(\theta+\alpha)=R \cos\theta,\\
 \label{eq.alphatheta}
\end{equation}
where $l=1$~m is the constant distance between the roller center
and the point at which the tape joins the winding spool. An
interesting limit case of Eq.~(\ref{eq.roller}) is then
obtained~\cite{De2004} when the roller radius $R$ is small
compared to the distance $l$, so that $\theta+\alpha\simeq \pi/2$.
In our experiments, it is almost the case, with $R/l<3\%$, and the
roller equation of motion (\ref{eq.roller}) can be approximated by
\begin{equation*}\nonumber
I \ddot{\beta} \simeq F R\,\sin\alpha. \tag{3b}\label{eq.roller2}
\end{equation*}
Then, assuming a uniform tensile strain in the peeled tape, the
force $F$ transmitted to the roller is simply
\setcounter{equation}{4}
\begin{equation}
F=\frac{E b e}{L-u}\,u, \label{eq.Fu}
\end{equation}
where $u$ is the elongation of the tape of Young modulus $E$,
thickness $e$ and width $b$. The assumption of a uniform peeled
tape strain amounts to neglect transverse waves in the tape under
tension. It is worth to note that these waves may however
influence the high frequency stick-slip instability in some
peeling regimes. In Eq.~(\ref{eq.Fu}), the peeled tape length $L$
is not a constant (see Fig.~\ref{fig:setup}) and varies with the
angle $\alpha$ according to
\begin{equation}
L(t)^2=l^2+R^2-2l R\cos\alpha(t).
\end{equation}
Experimentally, the observed instantaneous values of $\alpha$
range between $-25\degree$ and $+25\degree$ at most. Such
variations of $\alpha$ induce peeled tape length variations of
$\delta L/L \sim 0.3\%$ in our geometry. These very small
variations of $L$ during the peeling experiments should have no
significant impact on the velocity thresholds and the other
features of the stick-slip instability~\cite{Yamazaki2006}.

Finally, the following kinematical constraint on the peeled tape
elongation applies
\begin{equation}
V=v_p+\dot{u}-R\cos\theta\, \dot{\alpha}.
\label{eq.kin}
\end{equation}
Note the sign change in the last term of Eq.~(\ref{eq.kin})
compared to Ref.~\cite{De2004} due to the opposite orientation
chosen for $\alpha$. Using the approximation $\theta\simeq
\pi/2-\alpha$, Eq.~(\ref{eq.lp}) and the integration over time of
Eq.~(\ref{eq.kin}) give
\begin{equation*}
\ell_p-Vt=R(\alpha + \delta \beta) = u_0 - u+
R(\cos\alpha_0-\cos\alpha),\tag{7b}\label{eq.kin2}
\end{equation*}
in which $\delta \beta=\beta-V t/R$ measures the unsteady part of
the roller rotation. In Eq.~(\ref{eq.kin2}), $u_0$ and $\alpha_0$
are constants corresponding to the values of $u$ and $\alpha$ at
$t=0$ for which $\ell_p=0$ by definition. Then, since the peeling
crack length averaged over a long time $\langle \ell_p \rangle$
simply equals to $V t$, one gets $\langle u \rangle =
u_0+R(\cos\alpha_0-\langle \cos\alpha\rangle)$, where $\langle
\,\, \rangle$ denotes the time average, which measures the mean
level of deformation of the peeled tape during the experiment.

To close the system of equations describing the peeling
experiments, one needs to model how the peeling fracture velocity
$v_p$ is set. Such physical condition for peeling is usually
expressed as a balance between the elastic energy release rate $G$
of the system and the fracture energy $\Gamma$ required to peel a
unit surface such that
\begin{equation}
G=\Gamma(v_p).
\label{eq.peelphysics}
\end{equation}
$\Gamma(v_p)$ accounts for the energy cost of the dissipative
processes near the fracture front during the fracture growth. In
general, this fundamental quantity in fracture mechanics is
characteristic of the type of material to fracture, of the
fracture geometry and of the fracture velocity. For a given
material and geometry, it is therefore classically considered to
be a function of the fracture velocity $v_p$ only. In the context
of adhesive peeling, $\Gamma$ is therefore also characteristic of
the rheology of the adhesive material, of the backing and of the
substrate. Finally, it is a priori also a function of the local
geometry near the fracture front: the thickness of adhesive, the
local peeling angle... However, most theoretical works on
stick-slip adhesive peeling consider only the dependance of
fracture energy on fracture velocity $v_p(t)$, except in some
models which assume that $\Gamma$ is also dependent on the imposed
velocity $V$~\cite{De2004,De2005}.

The elastic energy release rate $G$ corresponds to the amount of
mechanical energy released by the growth of the fracture by a unit
surface. This quantity, which is geometry dependent, both takes
into account the work done by the operator and the changes in the
recoverable energy stored in material strains. The following
expression is traditionally used for the peeling fracture
geometry~\cite{Kendall1975,De2004}
\begin{equation}
G=\frac{F}{b}(1-\cos \theta).
\label{eq.G}
\end{equation}
This is a very good approximation for most adhesive tapes and
peeling geometries, except when the peeled tape stretching energy
cannot be neglected for very small peeling
angles~\cite{Kendall1975} or when its curvature elasticity has to
be taken into account~\cite{Kinloch1994} especially for very short
peeled tape length.

It is usually assumed that in the fracture propagation
equation~(\ref{eq.peelphysics}), the effect of peeling angle
$\theta$ is fully taken into account by its appearance in the
energy release rate~(\ref{eq.G}). In other words, it is usually
considered that $\Gamma$ itself does not depend on $\theta$.
Consequently, the velocity range in which stick-slip appears is
expected to be independent of the peeling angle and to be set
mainly by the region where $\Gamma(v_p)$ has a negative slope,
with some limitations due to an influence of the peeled tape
stiffness~\cite{Yamazaki2006}.

Altogether, we can identify three independent degrees of freedom
(for example $\alpha$, $\beta$ and $u$) related to each other by
the system of Eqs.~(\ref{eq.vp}-\ref{eq.G}) involving three
differential equations: (\ref{eq.roller}), (\ref{eq.kin}) and
(\ref{eq.peelphysics}). An interesting exact solution is the
steady state, or fixed-point, solution corresponding to a regular
peeling and given by
\begin{equation}
\begin{aligned}
\alpha&=0;\,\dot{\beta}=\frac{V}{R};\,\frac{u}{L}=\frac{1}{1+ E e/\Gamma(V)};\\
\theta &= \frac{\pi}{2};\,v_p=V;\,L=l-R;\,\frac{F}{b}=\Gamma(V).
\end{aligned}
\label{eq.steady}
\end{equation}

\section{Results}

\subsection{Basic Stick-Slip features}
\begin{figure}
\centerline{\includegraphics[width=7.8cm]{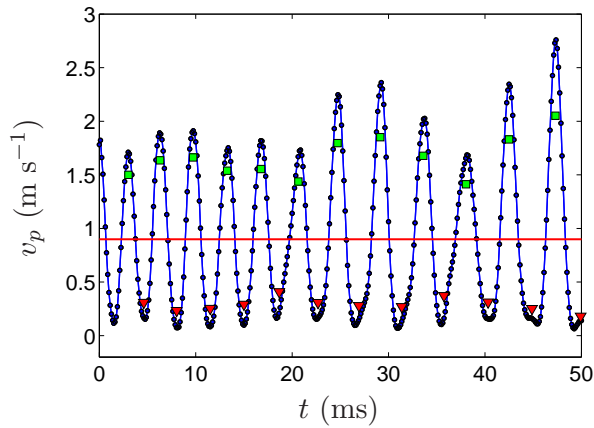}}
    \caption{(Color online) Peeling point velocity $v_{p}(t)$ in the roller reference
    frame for an experiment performed at $V=0.90$~m~s$^{-1}$.
    Triangles and squares respectively show the averaged stick
    $v_{stick}$ and slip $v_{slip}$ velocities for each stick-slip
    cycle. The horizontal straight line shows the imposed peeling
    velocity $V$.} \label{fig:oscillation}
\end{figure}

In Fig.~\ref{fig:oscillation}, we plot a typical signal of peeling
fracture velocity $v_p(t)$ for an imposed peeling velocity
$V=0.90$~m~s$^{-1}$. The observed large and oscillating
fluctuations of $v_p(t)$ are the characteristic signature of the
stick-slip motion. Note that the amplitude of these oscillations
is roughly as large as the mean peeling velocity. In particular,
the peeling experiences an almost complete arrest with a very low
fracture velocity (here, fluctuating between $0.05$~m~s$^{-1}\sim
0.06\,V$ and $0.15$~m~s$^{-1}\sim 0.17\,V$) once every stick-slip
cycle. The period of these oscillations is quite stable during an
experiment (here, $3.9\pm 0.4$~ms for $V=0.90$~m~s$^{-1}$).

Now considering all the experiments, over the whole range of peeling velocities
$0.25<V<2.45$~m~s$^{-1}$ for which we observe stick-slip
instability, the stick-slip oscillations period (averaged over all
the stick-slip events for each experiment) is very stable, in the
range $3.9\pm 0.3$~ms. This result is in contrast with the data
reported in~\cite{Barquins1986,Maugis1988} for a different
adhesive roller tape (3M Scotch$^\circledR$ 602) also peeled at
constant velocity. In~\cite{Barquins1986,Maugis1988}, the
stick-slip period was extracted from torque time series
provided by the winding motor and was indeed shown to be proportional to $L$ and
approximatively proportional to the inverse of $V$ over the whole
range of instable peeling velocities (which was
$0.06<V<2.1$~m~s$^{-1}$). The linearity of the stick-slip period
with $L/V$ reported in \cite{Barquins1986} agrees with a model
where the limit of stability of the stick phase, before the system
jumps into the slip phase, corresponds to the reach of a constant
threshold in strain or stress in the peeled ribbon. Indeed, during
the stick phase the peeled tape strain almost linearly increases
with time as $Vt/L$. An important assumption of the
model developed in~\cite{Barquins1986,Maugis1988} is that the slip
phase duration is negligible compared to the stick phase one.
However, in these works, this assumption remained untested since
the torque measurements did not allow a direct access
to the peeling fracture dynamics contrary to our measurements.
As can be seen in Fig.~\ref{fig:oscillation}, the assumption
of a negligible slip phase duration is obviously far from being true in our experiments which could
explain why this model fails here and also suggests that we are
not investigating a comparable stick-slip regime.

In our experiments, as a consequence of the constancy of the mean
stick-slip cycle duration $T_{ss}$, the mean amplitude of the
fracture propagation $A_{ss}$ during stick-slip cycles increases
almost proportionally to the peeling velocity $V$ according to
$A_{ss}= V\,T_{ss}$. It is however remarkable to note that the
dispersion inside a given experiment of the stick-slip cycles
amplitude and period is increasing significantly from about 5 to
40\% with the imposed velocity $V$ going from $0.25$ to
$2.45$~m~s$^{-1}$. We will see in the following that this
increasing dispersion is the trace of the growth with $V$ of low
frequency oscillations of the mean peeling angle (averaged over
one stick-slip event) which induce intermittencies in the
stick-slip instability.

From the signal of instantaneous peeling velocity, we actually
search for all the moments at which the sign of $v_p(t)-V$
changes. When $v_p(t)-V$ goes from positive to negative, it
defines the beginning of a stick event and when it goes from
negative to positive, it defines the beginning of a slip event. We
then compute the mean stick $v_{stick}$ and slip $v_{slip}$
velocities as the average value of the velocity $v_p(t)$ during
the phases where $v_p(t)<V$ (stick) and $v_p(t)>V$ (slip).
Finally, only the events during which $v_p$ is successively
smaller than $0.95\,V$ and larger than $1.05\,V$ are considered as
true stick-slip events. This allows to avoid measurement noise and
small velocity fluctuations to be taken into account as stick-slip
events during periods where no stick-slip is present. These stick
and slip velocities are reported in Fig.~\ref{fig:oscillation} as
triangle and square symbols respectively. We observe that the
stick and slip mean velocities are fluctuating in time during a
peeling experiment at constant velocity $V$. This is probably
mainly because of heterogeneities in the adhesion properties of
the peeled tape and also maybe, to a lesser extent, because of the
fluctuations of the imposed velocity.

\begin{figure}
\vspace{0.3cm} \centerline{\includegraphics[width=7cm]{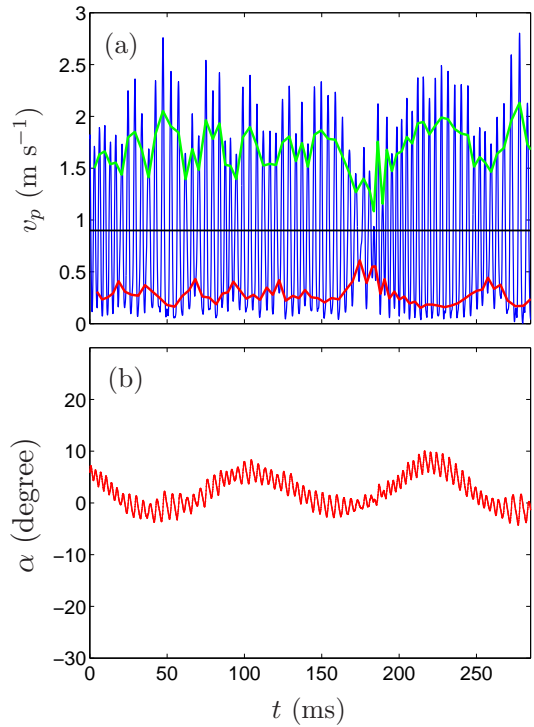}}
    \caption{(Color online) (a) Peeling point velocity $v_{p}$ in the roller
    reference frame as a function of time for an experiment performed
    at $V=0.90$~m~s$^{-1}$. The top and bottom continuous lines
    respectively trace the slip and stick local mean velocities. The
    horizontal straight line shows the average peeling velocity $V$.
    (b) shows the corresponding instantaneous peeling point angular
    position $\alpha$ as a function of
    time.}\label{fig:ppvelocity_osc}
\end{figure}

At the lower peeling velocities belonging to the instable
interval, the stick and slip velocities are however relatively
stable throughout the peeling cycles during an experiment as can
be seen in Fig.~\ref{fig:ppvelocity_osc}(a) (same experiment at
$V=0.90$~m~s$^{-1}$ as in Fig.~\ref{fig:oscillation}). We
nevertheless observe in this figure at time $t\sim 180$~ms that
the stick-slip amplitude decreases abruptly and temporarily during
three stick-slip cycles. We believe such ``accident'' may be
related to rare large scale defects in the adhesion of the
commercial tape.

\subsection{Stick-Slip intermittencies and roller pendular oscillations}

\begin{figure}
\centerline{\includegraphics[width=7cm]{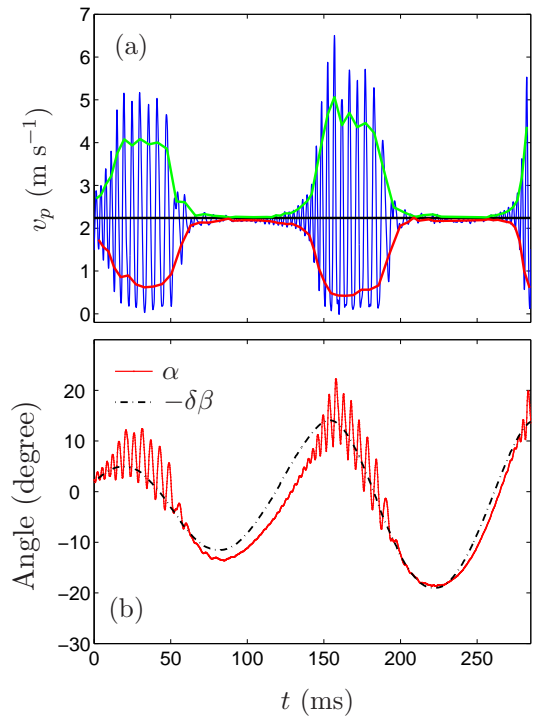}}
    \caption{(Color online) (a) Peeling point velocity $v_{p}$ in the roller
    reference frame and (b) angular positions $\alpha(t)$  and
    $-\delta \beta(t)\equiv Vt/R-\beta(t)$ as functions of
    time for an experiment performed at $V=2.24$~m~s$^{-1}$. Same
    layout as in Fig.~\ref{fig:ppvelocity_osc}.}\label{fig:ppvelocity_osc2}
\end{figure}

Remarkably, as the average peeling velocity $V$ is increased, we
observe that the stick-slip dynamics becomes intermittent,
alternating regularly between periods of time with fully-developed
stick-slip cycles and periods of time without or at least with
strongly attenuated stick-slip amplitude. A typical example of
such intermittencies is shown in Fig.~\ref{fig:ppvelocity_osc2}(a)
where a period of about 140~ms ($\sim 7$~Hz) can be seen.
Comparing these data with the instantaneous angular position of
the peeling point in the laboratory $\alpha(t)$ in
Fig.~\ref{fig:ppvelocity_osc2}(b), we see that the intermittent
stick-slip behavior is strongly correlated with low frequency
variations of this angle, whereas high frequency variations of
$\alpha(t)$ (at about $\sim 250$~Hz) are directly correlated to
the stick-slip motion.

The slow oscillations of the angular peeling position $\alpha(t)$
are the direct consequence of a low frequency pendulum-like motion
of the adhesive roller, in addition to its mean rotation at a rate
$V/R$. Indeed, as can be seen in
Fig.~\ref{fig:ppvelocity_osc2}(b), the angle $\delta
\beta(t)=\beta(t)-V \,t/R$, which measures the unsteady part of
the roller rotation, matches rather well the low frequency
oscillations of $-\alpha(t)$ when smoothing over the fast
stick-slip oscillations. This observation, $\langle\alpha +
\delta\beta\rangle_{ss} \simeq 0$, where $\langle \,\,
\rangle_{ss}$ stands for the average over a stick-slip cycle, can
be understood in the following way. Experimentally, we observe
that the mean (averaged over a stick-slip cycle) fracture velocity
$\langle v_p \rangle_{ss}$ is always equal to the imposed peeling
velocity $V$ to better than $7\%$. Therefore, to a good
approximation, we have $ \langle \ell_p \rangle_{ss} \simeq V t$.
Finally, using the first equality in Eq.~(\ref{eq.kin2}), this
shows that $\langle \alpha \rangle_{ss} \simeq - \langle \delta
\beta \rangle_{ss}$ as is indeed verified in
Fig.~\ref{fig:ppvelocity_osc2}(b). Furthermore, averaging
Eq.~(\ref{eq.roller2}) over a stick-slip cycle and using $\langle
\alpha \rangle_{ss} \simeq - \langle \delta \beta \rangle_{ss}$,
we get
\begin{equation}\label{eq:sloosc}
\langle \ddot{\delta \beta}\rangle_{ss}+\frac{FR}{I}\langle\sin\delta \beta \rangle_{ss} \simeq 0,
\end{equation}
which predicts pendular oscillations of the unsteady part of the
roller rotation at a frequency close to $\omega=\sqrt{FR/I}$ for
small amplitudes of $\delta \beta$.

To check this interpretation of the pendular oscillations, we have
made some measurements of the mean peeling force $\langle F
\rangle$, time averaged over the whole constant velocity peeling
experiment. This is done with a force gage (Interface$^\circledR$
SML-5), aligned with the direction $\alpha=0$, and placed between
the adhesive roller pulley and its mechanical support. In
table~\ref{tab.freq}, we compare the frequency of the slow
oscillations with the characteristic frequency
$\omega=\sqrt{\langle F \rangle R/I}$ replacing $F$ by its
temporal average value. Although this framework is only
approximate, we find a rather good agreement between the direct
measurement of the period and the theoretical prediction
$2\pi/\omega$. We conclude that the low frequency dynamics
develops due to the interplay between the inertia of the roller
and the moment applied to the roller by the peeling force as
already suggested in \cite{Cortet2007}.
\begin{table}
\begin{tabular}{|c|c|c|c|c|}
\hline
 $V$ (m.s$^{-1}$)& $\langle F \rangle$ (N) &\,$T\,(s)$\, & \,$2\pi/\omega\,(s)$\,  \\ \hline
 0.36 $\pm$ 0.01 & 1.71 $\pm$ 0.07 & 0.109 $\pm$ 0.005 & 0.092 $\pm$ 0.002  \\ \hline
 0.50 $\pm$ 0.01 & 1.40 $\pm$ 0.06 & 0.115 $\pm$ 0.005 & 0.102 $\pm$ 0.002 \\ \hline
 0.72 $\pm$ 0.02 & 1.18 $\pm$ 0.05 & 0.118 $\pm$ 0.005 & 0.111 $\pm$ 0.002 \\ \hline
 1.53 $\pm$ 0.03 & 0.91 $\pm$ 0.04 & 0.130 $\pm$ 0.005 & 0.126 $\pm$ 0.003\\ \hline
\end{tabular}
\caption{Comparison between the direct measurement of the low
frequency oscillations period $T$ and the period
$2\pi/\omega=2\pi/\sqrt{\langle F \rangle R/I}$ estimated using
the average peeling force $\langle F \rangle$ in
Eq.~(\ref{eq:sloosc}).} \label{tab.freq}
\end{table}

In the two previous paragraphs, we have shown that the slow
pendular oscillations of the adhesive roller are independent of
the physics of the adhesive fracture propagation. We have indeed
verified that the roller rotation $\beta(t)=Vt/R+\delta \beta(t)$
is unsensitive to the high frequency stick-slip oscillations of
$\alpha(t)$ and $v_p(t)$ because of the roller inertia.
Consequently, we feel entitled in the following to consider the
slowly oscillating mean peeling angle $\langle \theta \rangle_{ss}
\simeq \pi/2-\langle \alpha \rangle_{ss}\simeq \pi/2+\langle
\delta \beta \rangle_{ss}$ as an effective control parameter for
the fracture problem (i.e. Eq.~(\ref{eq.peelphysics})), which is
quasi-statically varying.

\begin{figure}
\centerline{\includegraphics[width=8cm]{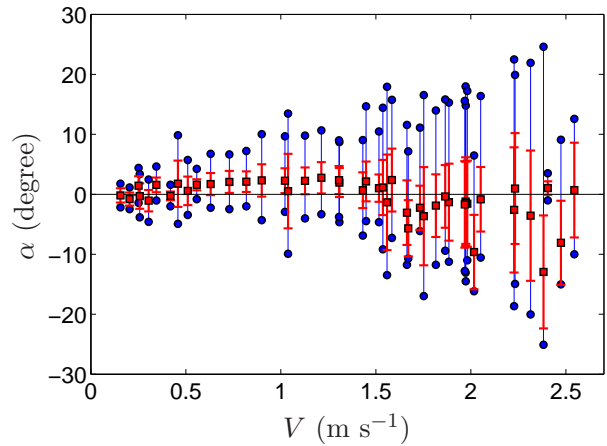}}
    \caption{(Color online) Mean angle $\alpha$ (squares) during each experiment and
    the corresponding standard deviation of its oscillations as
    errorbars. Circles show the maximum and minimum angle $\alpha$
    reached during each experiment.} \label{fig:alpha}
\end{figure}

In order to quantify the slow oscillations of the peeling point
angular position for various imposed velocity $V$, we plot as a
function of $V$ the mean angle $\alpha$ during each experiment and
the corresponding standard deviation of its oscillations as
errorbars (Fig.~\ref{fig:alpha}). We also report the maximum and
minimum angle $\alpha$ reached during each experiment. We can note
the regular increase of the oscillation amplitude of $\alpha$ from
$\sim \pm 2\degree$ up to $\sim \pm 25\degree$ as the imposed
velocity increases in the instable range, whereas its mean value
is quite stable in the range $\alpha\in[-4, 3]\degree$. Since the
effective peeling angle verifies $\theta\simeq\pi/2-\alpha$, it
has a mean value always close to $\theta\simeq 90\degree$,
corresponding to the steady state solution~(\ref{eq.steady}), and
variations up to $\pm 25\degree$ around the mean at large peeling
velocities.

In Fig.~\ref{fig:ppvelocity_osc2}, we see that large amplitude
stick-slip occurs mostly for the larger and positive values of
$\alpha(t)$ (i.e., $\theta< 90\degree$) whereas for negative
values (i.e., $\theta> 90\degree$), stick-slip almost disappears.
Such straightforward correlation is however a simplistic picture
since it can also be noted that there is some hysteresis in the
angle $\alpha$ at which stick-slip appears and disappears. Guesses
could be that the hysteresis is due to a delayed response of the
peeling instability when the angle $\alpha$ changes, which would
corresponds to a value of the stick-slip instability growth rate
comparable to the pendular oscillations frequency. More generally,
this hysteresis may reveal dynamical effects related to
$d\theta/dt$. At low peeling velocity
(Fig.~\ref{fig:ppvelocity_osc}(b)), low frequency oscillations of
the peeling point angle do actually already exist but, as we have
seen, are of smaller amplitude. They moreover apparently do not
correlate with small stick-slip amplitude modulations. This
suggests that the slow oscillations of $\alpha$ must overtake a
certain amplitude to trigger a significant time modulation
of the stick-slip amplitude.

\subsection{Stick and Slip velocities, and correlation with peeling angle}

\begin{figure}
\centerline{\includegraphics[width=7.5cm]{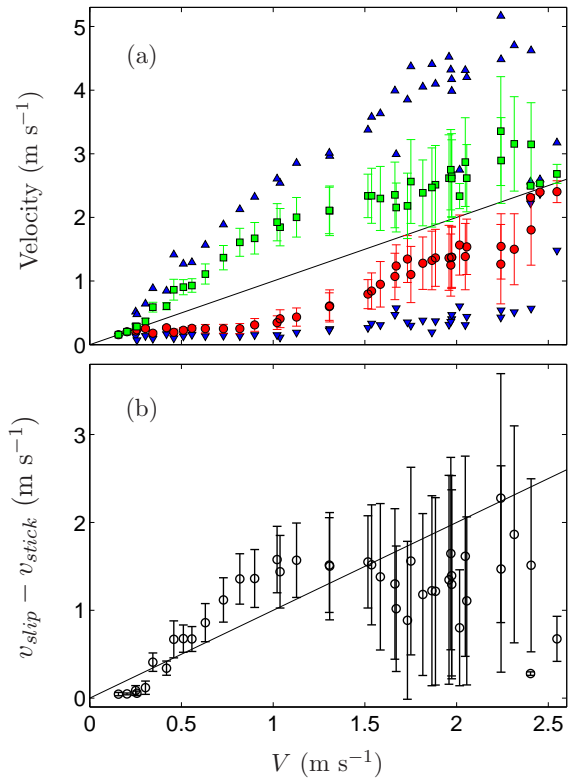}}
    \caption{(Color online) (a) Average slip (squares) and stick (circles) velocities
    and maximum slip (up triangle) and minimum stick (down triangle)
    velocities and (b) average of the difference $v_{slip}-v_{stick}$,
    as a function of the imposed peeling velocity $V$. In (a) and (b),
    the continuous line corresponds to the imposed peeling velocity.
    Each data point is an average and the error bar the standard
    deviation over all stick-slip events in a single experiment. The
    large values of standard deviation at large peeling velocities are
    the trace of the intermittent occurrence of stick-slip.}
    \label{vit_stick_slip1}
\end{figure}

In Fig.~\ref{vit_stick_slip1}(a), we plot the average (over all
the events in each experiment) stick and slip velocities as a
function of the imposed peeling velocity $V$. For the lower
peeling velocities, we have plotted $v_{stick}=v_{slip}$ which
means that the peeling is regular without observation of
stick-slip events. The stick-slip actually initiates at a peeling
velocity threshold of $0.25\pm 0.02$~m~s$^{-1}$ with average stick
and slip velocities starting to deviate from the imposed peeling
velocity $V$ (continuous line). This threshold corresponds very
well to the value measured for the same roller adhesive tape
peeled by falling loads~\cite{Cortet2007}. The stick and slip
velocities increase gradually for $V$ varying from $0.25$ up to
$2.45\pm 0.10$~m~s$^{-1}$ for which value they collapse on the
average velocity $V$. The measured disappearance threshold for
stick-slip at large velocities, $2.45\pm 0.10$~m~s$^{-1}$, is also
compatible with the previously measured value in peeling
experiments by falling loads where it was about $2.6$~m~s$^{-1}$.

In Fig.~\ref{vit_stick_slip1}(a), the data are accompanied with
their corresponding statistical standard deviation inside each
experiment. These standard deviations are quite low ($\sim 5$ to
$10\%$) from $V=0.25$ to $1.5$~m~s$^{-1}$ which means that the
corresponding stick-slip features are quite stable during a given
experiment. For average velocities $V$ larger than
$1.5$~m~s$^{-1}$ and up to the disappearance of the stick-slip at
$2.45\pm 0.10$~m~s$^{-1}$, we observe larger standard deviations
($\sim 10$ to $20\%$) for the stick and slip velocities. This
increase is obviously the trace of the stick-slip intermittencies
that lead to alternate periods of strong and weak stick-slip
oscillations.

Finally, in Fig.~\ref{vit_stick_slip1}(a), we also plot the
maximum slip and minimum stick velocities measured during each
experiment. We see that as the peeling becomes more and more
intermittent with the increasing peeling velocity $V$, the extreme
values of the stick and slip velocities are further and further
away from the average ones which reveals the amplitude of the
stick-slip modulations. Focusing on the two experiments at imposed
velocity $V=2.40$~m~s$^{-1}$, we can observe one experiment with a
developed stick-slip and one experiment with almost no remaining
stick-slip with mean stick and slip velocities about only $4\%$
smaller and larger than $V$ respectively. These observations
reveal the unprecise definition of the stick-slip disappearance
threshold which is an intrinsic feature of adhesive stick-slip,
amplified in the present case by the slow oscillations of the
peeling angle. Regarding the mean velocities, the last two data
points, at $2.47$ and $2.55$~m~s$^{-1}$, show an almost complete
absence of stick-slip. On the contrary, we see that the maximum
slip and minimum stick velocities are very close to $V$ for
$2.47$~m~s$^{-1}$ but quite far anew for the experiment at
$2.55$~m~s$^{-1}$: in the last case, this is simply the trace of
very marginal stick-slip events existing only during short phases
of the pendular oscillations where the angle $\alpha(t)$ is large.

To study these intermittencies in more details, in
Fig.~\ref{vit_stick_slip1}(b), we plot as a function of the
imposed velocity the quantity $v_{slip}-v_{stick}$ averaged over
all stick-slip cycles in each experiment. We see that the mean
velocity amplitude of stick-slip is first larger than the imposed
velocity up to $V=1.5$~m~s$^{-1}$ before being overall lower and
quite scattered as a consequence of the stick-slip
intermittencies. Here, again the errorbars correspond to the
standard deviation of the plotted statistical quantity. This data
illustrates very well the strong increase of the explorated range
of stick-slip amplitudes as the peeling velocity $V$ increases.
One can indeed observe in Fig.~\ref{vit_stick_slip1}(b) that the
standard deviation of the stick-slip amplitude becomes almost as
large as its mean value for $V>1.7$~m~s$^{-1}$ which is the trace
of the strongly intermittent behavior.

\begin{figure}
\centerline{\includegraphics[width=8.5cm]{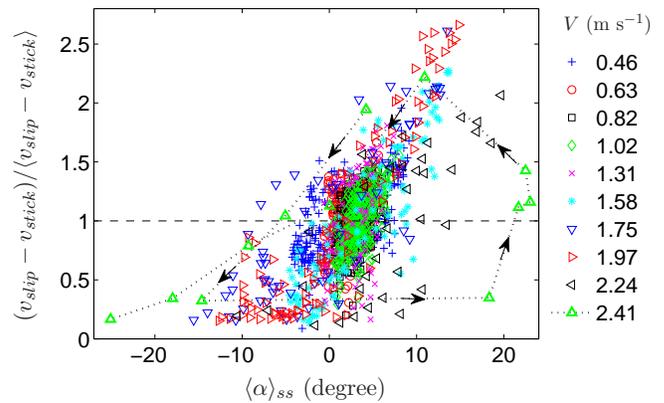}}
    \caption{(Color online) Parameter, $(v_{slip}-v_{stick})/\langle
    v_{slip}-v_{stick}\rangle$, quantifying the normalized dependance
    of the velocity contrast between the slip and stick phases with
    the angular position of the peeling point $\langle \alpha
    \rangle_{ss}$ for various imposed peeling velocities $V$. Each
    data point corresponds to a single stick-slip event.
    The dotted line and the arrows indicate the time sequence
    of successive stick-slip events in the $V=2.41$~m~s$^{-1}$
    experiment which reveals a large hysteresis loop.}
\label{fig:param}
\end{figure}

Finally, in order to quantify the correlations between the peeling
point angular position and the amplitude of stick-slip, we
introduce an order parameter defined as the difference between the
slip and stick velocities for each stick-slip event,
$(v_{slip}-v_{stick})/\langle v_{slip}-v_{stick}\rangle$,
normalized by its average over all the events at a given imposed
velocity. Fig.~\ref{fig:param} shows the evolution of this
parameter as a function of the mean angular position of the
peeling point $\langle \alpha \rangle_{ss}$ for each stick-slip
cycle during the experiments and for a wide selection of imposed
velocity $V$. We first see that the average operating point in
each data series at a given imposed velocity $V$, which is defined
by $v_{slip}-v_{stick}=\langle v_{slip}-v_{stick}\rangle$,
corresponds for a large majority of events to angles in the region
$\langle \alpha \rangle_{ss}\in[0\degree,5\degree]$. This
observation is the trace of the fact that, without the parasitic
pendular oscillations of the roller which generate the
intermittencies, the stick-slip peeling would naturally proceed
with a mean peeling angle in the range
$\langle \theta \rangle_{ss}\in[85\degree,90\degree]$. Around this operating
point $(v_{slip}-v_{stick}=\langle
v_{slip}-v_{stick}\rangle$,\,$\langle \alpha
\rangle_{ss}\in[0\degree,5\degree]$), the statistics of the
stick-slip events gather on a cloud which can be (roughly)
modelled by
$$v_{slip}-v_{stick}=g(V) \times f(\langle \alpha \rangle_{ss}),$$
with $f$ a rapidly increasing function and a separation of the
variables $V$ and $\langle \alpha \rangle_{ss}$. Here, $g$ is
defined as the mean velocity contrast, $g(V)=\langle
v_{slip}-v_{stick}\rangle(V,\alpha=\alpha_0)$, for a given stable
peeling angle $\alpha_0$. These data confirm that the stick-slip
instability increases dramatically in amplitude with $\langle
\alpha \rangle_{ss}$ and occurs preferentially when $\langle
\alpha \rangle_{ss}>-5\degree$ whereas it tends to disappear when
$\langle \alpha \rangle_{ss}<-5\degree$. These results overall
point out an important effect of the peeling angle $\theta\simeq
\pi/2-\alpha$ (Fig.~\ref{fig:setup}) on the stick-slip instability
thresholds and amplitude.

Speaking more accurately, the order parameter
$(v_{slip}-v_{stick})/\langle v_{slip}-v_{stick}\rangle$
dependence as a function of the angle $\langle \alpha
\rangle_{ss}$ does obviously not collapse perfectly on a master
curve $f$ in Fig.~\ref{fig:param}. It actually shows an hysteresis
that becomes stronger at large velocities (see the arrows
indicating the time sequence of successive stick-slip events
in the $V=2.41$~m~s$^{-1}$ experiment). As already mentioned, we attribute this
hysteresis to a delay in the response of the peeling instability
to a change in the experimental peeling angle $\theta$ or to the
dynamical effects of $d\theta/dt$. Nevertheless, this hysteresis
is far beyond our current understanding of the adhesive stick-slip
peeling. To the first order, we therefore believe that this
overall dependance of the stick-slip amplitude with the local mean
(over each stick-slip cycle) peeling angle $\langle \theta(t)
\rangle_{ss}$ reflects a general intrinsic dependance of the
peeling fracture process with the peeling angle $\theta$, which
should be explored in peeling experiments at imposed mean angle
$\langle \theta \rangle_{ss}$.

\section{Discussion}
Theoretically, the angle $\theta$ at which the peeling of an
adhesive tape is performed is usually taken into account in the
calculation of the elastic energy release rate $G$ through
Eq.~(\ref{eq.G}). If one further assumes that the fracture energy
$\Gamma(v_p)$ is independent of the peeling angle as suggested by
Kendall's experiments in the regular peeling regime, the velocity
thresholds for the onset of stick-slip instability, related to the
zone where $\Gamma(v_p)$ is a decreasing function, should also be
roughly independent of the effective peeling angle $\theta$. In
that case, there are consequently no clear reasons for stick-slip
to be strongly dependent on the peeling angle at a given mean
fracture velocity $\langle v_p \rangle_{ss}=V$ in the instable
range of $\Gamma(v_p)$. The susceptibility of the stick-slip
instability to the peeling angle that we report in this article
therefore questions which are the correct dissipation mechanisms
that should be taken into account in the fracture energy $\Gamma$
during the instable regime of the peeling.

The behavior we have observed in Fig.~(\ref{fig:ppvelocity_osc2})
resembles to some extent the dynamics predicted by some models
(see for instance  Fig.~4(b) in \cite{De2004}). Here, the authors
have assumed that the fracture energy is a function of both the
local peeling velocity $v_p$ and the imposed velocity $V$ so that
$\Gamma(v_p,V)$, which can be viewed as an \textit{ad hoc} guess.
In the roller geometry, this model sometimes predicts a stick-slip
dynamics corresponding to high frequency oscillations of the angle
$\alpha$ superimposed to a lower frequency and larger amplitude
variation. The authors explain that this behavior is obtained
either when increasing peeling velocity for a given inertia of the
roller or when increasing the roller inertia for a given peeling
velocity. Thus, the intermittent appearance and disappearance of
stick-slip observed in this model seems to be the consequence of a
subtle balance between the effect of inertia of the roller and the
effect of a fracture energy depending explicitly on both the
pulling velocity $V$ and the fracture velocity $v_p$.

Another possibility to understand the observed stick-slip dynamics
would be that the fracture energy itself depends on the peeling
angle $\theta$ so that $\Gamma(v_p,\theta)$. From static
equilibrium considerations, it is clear that varying the angle of
peeling will change the relative contribution of normal and shear
load on the adhesive at the peeling front. Since it has been
observed that shear can have an effect on the resistance of
adhesives to rupture~\cite{Amouroux2001}, one could think that it
can also have an effect on the dependence of the fracture energy
with velocity, contrary to the results of
Kendall~\cite{Kendall1975}. The onset of stick-slip instability
would then naturally become dependent on the peeling angle.

At this point, it is not possible to conclude whether the
intermittent stick-slip behavior observed in our experiments is
due to inertial effects of the roller combined with a
$\Gamma(v_p,V)$ dependence of the fracture energy as proposed
in~\cite{De2004}, or if it is rather due to a direct dependence
$\Gamma(v_p,\theta)$ with the angle. Experiments performed in a
different geometry, such as peeling from a flat surface at
constant angle $\theta$, would help distinguish between the two
proposed mechanisms.

\begin{acknowledgments}

This work has been supported by the French ANR through grant
``STICKSLIP'' No. 12-BS09-014-01.

\end{acknowledgments}

\end{document}